\documentclass[
twocolumn,
longbibliography,
showpacs,                     
showkeys,                  
secnumarabic,
amssymb, 
nobibnotes, 
floatfix,
aps, 
pre]{revtex4-1}

\usepackage{graphicx}
\usepackage{latexsym}
\usepackage{amsmath} 
\usepackage{verbatim}

\begin{document}

\title{Comparison of canonical and microcanonical definitions of entropy}

\author{Michael Matty}
\author{Lachlan Lancaster}
\author{William Griffin}

\author{Robert H. Swendsen}
\email[]{swendsen@cmu.edu}
\affiliation{Department of Physics, Carnegie Mellon University, Pittsburgh PA, 15213, USA}

                              
\date{\today}

\begin{abstract}
For more than 100 years, one of the central concepts in statistical mechanics has been the microcanonical ensemble, 
which provides a way of calculating the thermodynamic entropy for a specified energy.  
A controversy has recently emerged between two distinct definitions of the entropy based on the microcanonical ensemble: 
(1) The Boltzmann entropy, defined by the density of states at a specified energy, 
and (2) The Gibbs entropy, defined by the sum or integral of the density of states below a specified energy.  
A critical difference between the consequences of these definitions pertains to the concept of negative temperatures, 
which by the Gibbs definition cannot exist.  
In this paper, we call into question the fundamental assumption that the microcanonical ensemble should be used to define the entropy.  
We base our analysis  on a recently proposed canonical definition of the entropy as a function of energy.  
We investigate the predictions of the Boltzmann, Gibbs, and canonical definitions for a variety of classical and quantum models.
Our results support the validity of the concept of negative temperature, but not for all models with a decreasing density of states. 
We find that only the canonical entropy consistently predicts the correct thermodynamic properties, 
while microcanonical definitions of entropy, including those of Boltzmann and Gibbs, are correct only for a limited  set of  models.
For models which exhibit a first-order phase transition,
we show that the use of the thermodynamic limit,
as usually interpreted, 
can conceal the essential physics.
\end{abstract}

\maketitle

\section{Introduction}
\label{introduction}

The thermodynamic entropy
is unique,
and 
provides a complete description of the 
macroscopic properties of a system\cite{lieb1997guide}.
Nevertheless,
its definition is still the subject of some dispute.
Recently,  a controversy has emerged between 
two distinct microcanonical definitions:
(1) The Boltzmann entropy, given by the density of states at a specified energy\cite{Boltzmann,Boltzmann_translation,Boltzmann_entropy,Planck_1901,Planck_book}, 
and (2) The Gibbs entropy, given by the sum or integral of the density of states below a specified energy\cite{Gibbs,Gibbs_entropy}. 
A critical difference between the consequences of these definitions pertains to the concept of 
negative temperatures\cite{Purcell_Pound,Ramsey_Neg_T,Landsberg_1959},
which by the Gibbs definition, cannot exist.  
The advocates of the Gibbs entropy
reject negative temperatures, claiming that they are inconsistent with thermodynamic principles
\cite{DH_Physica_A_2006,Romero-Rochin_2013,DH_NatPhys_2014,HHD_2014,Campisi_SHPMP_2015,HHD_2015,Sokolov_2014},
while
other authors have argued
that thermodynamics 
 is consistent with negative temperatures,
 and a different definition of entropy 
 can give correct thermodynamic predictions 
when the Gibbs entropy does not\cite{RHS_1,RHS_4,RHS_5,RHS_8,RHS_9,RHS_unnormalized,Vilar_Rubi_2014,Schneider_et_al,Frenkel_Warren_2015,DH_reply_to_FW,SW_2015_PR_E_R,SW_2016_Physica,DV_Anghel_2015,Cerino_2015,Poulter_2015,RHS_continuous}.

A related issue that has been  raised
is whether the limit of an infinite system
(thermodynamic limit)
 is essential
to  thermodynamics\cite{HHD_2014,Campisi_SHPMP_2015,HHD_2015,Touchette_introduction_2004,PhysRevE.74.010105,Touchette_equivalence_2011,Touchette_general_equivalence_2011}.
We take the position that,
while the approximation of an infinite system can be useful for certain calculations,
it is also necessary for any theoretical approach to specify how to 
 calculate the thermal properties of 
finite systems.

For example,
a gas in a container that can adsorb particles on its walls
has both interesting (non-extensive) physics and practical applications.
However,
in the limit of an infinite system 
the contribution of the walls diverges more
slowly than the contributions of the bulk.
As the thermodynamic limit is usually represented,
 the system appears to be extensive,
and the effect of the walls is lost.
We will  
show that in a similar way,
the thermodynamic limit can
 obscure the essential physics of 
first-order transitions,
and we suggest an alternate representation
of the thermodynamic limit.

It has been claimed that thermodynamics should  also apply 
to systems as small as a single 
particle\cite{DH_Physica_A_2006,Romero-Rochin_2013,DH_NatPhys_2014,HHD_2014,Campisi_SHPMP_2015,HHD_2015}.
While we agree that thermodynamics should apply to finite systems,
the predicted measurements of such systems should be unique. 
 This requires 
 a large number of particles
so  that the 
 measured macroscopic variables 
have  relative fluctuations
 smaller than the accuracy of the measurement.

The Boltzmann entropy predicts that
negative temperatures should occur
wherever the density of states is a decreasing function of energy.
Since this often occurs in a quantum spin system,
the entropy in quantum statistical mechanics is  central to the debate.
Defining the entropy for a quantum system 
has the added complexity that 
energy eigenvalues for a finite system 
are restricted to a discrete set of energies.
It has been recently pointed out that while 
microcanonical proposals for the entropy
are
``\emph{a priori}
only defined on the discrete set of eigenvalues''\cite{HHD_2014},
the correct thermodynamic  entropy,
even for quantum systems,
must be a continuous function of energy\cite{RHS_continuous}.
The key point in this argument is that if a system of interest has ever been in thermal contact 
with another system,
 separation will never leave either system in a quantum eigenstate.
Consequently,
the microcanonical ensemble is not an appropriate tool  
for calculating the thermodynamic properties of a quantum system.

It was also argued in Ref.~\cite{RHS_continuous}
that the appropriate probability distribution of quantum systems 
should be the canonical distribution,
even if a system of interest were to be separated from a smaller system.
This argument has been independently confirmed
 by explicit computations\cite{Jin_2013_1,Jin_2013_2,Novotny_decoherence_2015,Novotny_2016_1}.
 The consequence is that 
 the fundamental relation
 $S=S(U,V,N)$
 should be calculated with the canonical ensemble,
 as described in Ref.~\cite{RHS_continuous}.
 Because of the central role of the canonical ensemble,
 we refer to 
 this expression  
 as  the canonical entropy.
 Its essential features  
 are described 
in Subsection~\ref{subsection: continuous entropy}.

Even though the quantum microcanonical ensemble fails to provide 
an expression for the entropy as a continuous function of energy
for finite systems,
this still leaves the possibility that it could give a correct expression 
for the entropy 
in the limit of an infinite system,
which might then be an acceptable approximation for large systems.
In this paper,
we test the predictions of both microcanonical and canonical expressions 
for the entropy for a variety of models.
Our results show that 
although a microcanonical calculation can 
give the correct thermodynamic entropy 
for models 
with a monotonic density of states
in the limit of large systems,
its usefulness for more general problems 
in statistical mechanics is  restricted.

The  argument against using the quantum microcanonical ensemble 
also applies to the classical case.
If two macroscopic classical systems are separated,
the energies of each system will not be known exactly,
even if the total energy were (somehow) known. 
Each system will be described by a canonical ensemble,
from which 
 $S=S(U,V,N)$
and all other thermodynamic properties can be calculated.

We begin in Section~\ref{section: quantum without PT}
with quantum models 
that do not exhibit a first-order phase transition
to show the basic finite-size effects.
This includes 
 two-level systems,
the Ising model,
and
simple harmonic oscillators.

In Section \ref{section: first order},
we discuss first-order transitions,
beginning with the twelve-state Potts model in two dimensions.
With data taken 
from a Wang-Landau simulation\cite{Wang_Landau},
we find that the 
microcanonical results are incorrect,
while the canonical entropy 
correctly describes the thermodynamic behavior.
We discuss the thermodynamic limit and show
 that it  obscures  the 
physics of first-order phase transitions,
unless the limit is taken as shown in Fig.~\ref{Potts_16_32_rescale}.
The canonical entropy is shown to give the correct thermodynamic behavior
for both finite and infinite systems.

Finally, 
in Section~\ref{section: classical models}
we investigate a number of generic models with 
continuous densities of states.
Such models must be regarded as being classical,
since all finite quantum systems 
have discrete energy spectra.
Although we find that the microcanonical entropies 
correctly describe some models,
they are not reliable for models 
for which the density of states is not monotonically increasing.

In the next section,
we review the two proposed microcanonical  definitions of entropy,
along with the canonical entropy.

\section{Competing definitions of quantum entropy}
\label{section: definitions}

The first subsection 
reviews the microcanonical definitions of entropy 
for quantum systems,
and
the second subsection
reviews the calculation of the canonical entropy 
as a continuous function of energy.

\subsection{Definitions of microcanonical entropy}
\label{subsection: microcanonical S}

We  assume that the quantum eigenvalue problem for a system of interest has been solved,
so that we know the spectrum of energy eigenvalues
$\{ E_n \}$ for quantum numbers $n$,
along with the degeneracies of the $n$-th energy level,
$\omega( E_n )$.
The Boltzmann entropy is then defined as 
\begin{equation}\label{definition SB}
S_B = k_B \ln \omega ( E_n )  .
\end{equation}
For the Gibbs entropy, 
we introduce  the notation 
\begin{equation}\label{definition Omega}
\Omega( E_n ) = \sum_{  E_m \le E_n}  \omega ( E_m )  ,
\end{equation}
to define 
\begin{equation}\label{definition SG}
S_G = k_B \ln \Omega ( E_n )  .
\end{equation}
Both $S_B$ and $S_G$ 
are defined only on the discrete set of energy eigenvalues
for a quantum system.

For classical systems,
$\omega(E)$ 
is defined by an integral over a constant energy surface in phase space,
while
$\Omega(E)$ 
is defined by the volume enclosed by that surface.

\subsection{Definition of  the canonical entropy }
\label{subsection: continuous entropy}

The calculation of the thermodynamic entropy 
proposed in 
Ref.~\cite{RHS_continuous}
proceeds through the canonical ensemble,
so we  refer to it as the canonical entropy.
As mentioned in the introduction,
any macroscopic system that had ever been in thermal contact with 
another macroscopic system of any size
will not be in an energy eigenstate,
and should be described by 
a canonical ensemble\cite{RHS_continuous,Novotny_decoherence_2015}.

The thermodynamic energy at a given temperature $T$
can be obtained from the usual equation
\begin{equation}\label{U 1}
U 
=
\frac{1}{Z}
 \sum_{n }
  E_n \,
  \omega( E_n )
\exp \left( -
\beta
  E_n
  \right)          ,
\end{equation}
where $\beta = 1/k_B T$, 
and the partition function is given by
\begin{equation}\label{Z 1}
Z 
=
 \sum_{n }
  \omega( E_n )
\exp \left( -
\beta
  E_n
  \right)            .
\end{equation}

Since we are going to analyze systems for which the canonical entropy,
$S=S(U)$,
is not necessarily a monotonic function of 
$U$,
it  is convenient to use a Massieu function,
rather than the more usual Helmholtz free energy,
$F=U[T]$,
where the square brackets around 
$T$ 
indicate the Legendre transform\cite{Callen,RHS_book}.
We define a dimensionless entropy 
$\tilde{S}_C(U)=S_C(U)/k_B$,
which has the differential form 
\begin{equation}
d \tilde{S} = \beta \, dU + \beta P dV - \beta \mu \, dN .
\end{equation}
Now perform a Legendre transform
with respect to
\begin{equation}\label{beta=dS/dU}          
\beta
=
\left(
\frac{ \partial \tilde{S} }{ \partial U }   
\right)_{V,N}  .
\end{equation}
 to obtain 
the Massieu function
\begin{equation}\label{S[b]=S-bU}                   
\tilde{S}[\beta] = \tilde{S} - \beta U = - \beta \left( U - TS \right) = - \beta F ,
\end{equation}
where the square brackets
indicate that 
$\tilde{S}[\beta]$
is the Legendre transform of 
$\tilde{S}$
with respect to
$\beta$.
From 
Eq.~(\ref{S[b]=S-bU})
and the well-known relation
$Z= \exp ( - \beta F )$,
we find the simple expression
$\tilde{S}[\beta]  = \ln Z$.

To obtain the canonical entropy, we need only perform 
an inverse Legendre transform to find
\begin{equation}\label{St = St b + b U}
\tilde{S}_C = \tilde{S}[\beta] +\beta(U) U .
\end{equation}
Finally, the entropy with the usual dimensions is given by 
$S_C = k_B \tilde{S}_C$  .

An important feature of the canonical entropy 
is that it can be used to calculate equilibrium conditions between 
macroscopic systems 
with incommensurate energy level spacings.

We begin our comparisons of the consequences of these
definitions with  relatively simple cases 
of models that do not exhibit phase transitions.

\section{Quantum models without a 
first-order phase transition}
\label{section: quantum without PT}

In earlier work,
$S_B$, $S_G$, and $S_C$
were calculated and compared for 
independent two-level objects
and 
independent simple harmonic oscillators\cite{RHS_continuous}.
The canonical entropy is strictly extensive
for these models,
as might be expected for independent objects.
However,
neither of the micocanonical definitions
has this  property.

In the following 
subsection,
we extend these calculations to the 
two-dimensional Ising model
of interacting spins.
We then review the results 
for simple harmonic oscillators.

\subsection{Two-dimensional Ising model}
\label{subsection: Ising model}

\begin{figure}[htbp]
\begin{center}
 \includegraphics[width=\columnwidth]{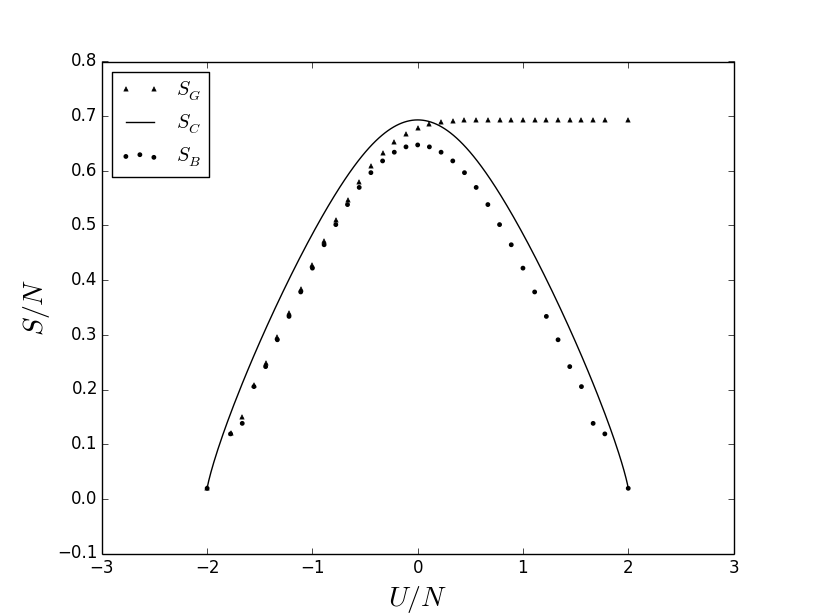}\\
  \caption{The solid line is a plot of the
  canonical entropy 
  $S_C$ 
  of a 
  $L \times L$ Ising model with $L=6$
  as a function of the thermodynamic energy $U$.
  The circles show $S_B$ and the triangles show $S_G$.
    All data were calculated from the exact degeneracies of the 
    energy levels,
    $\omega(E_n)$, which were obtained with the Beale 
    algorithm\cite{Beale_1996}.
    }
  \label{Ising_plot}
\end{center}
\end{figure}

The Ising model consists of a set of spins 
$\sigma_j= \pm 1$ occupying sites on a lattice.
We  consider a two-dimensional model on a square lattice
of size $L \times L$,
with periodic boundary conditions.
The Hamiltonian is given by 
\begin{equation}
H_I = - J \sum_{\langle j,k \rangle}  \sigma_j  \sigma_k   ,
\end{equation}
where
$J$ is an interaction energy,
and
 the sum is over nearest-neighbor pairs of sites,
$\langle j,k \rangle$.
An algorithm for the exact determination of the 
degeneracies of the energy levels,
$\omega(E_N)$,
has been given by Beale\cite{Beale_1996}.
Using the exact values of $\omega(E_N)$,
we have computed the canonical entropy 
for various lattice sizes and compared the resulting functions
with 
$S_G$ and $S_B$.
Fig.~\ref{Ising_plot} 
shows the results 
for $L=6$.
%


It can be seen that the solid curve representing the canonical entropy
$S_C$
lies above $S_B$ for all energies 
except the highest and lowest,
for which they both take on the expected value of 
$k_B \ln 2$.
While 
$S_G$ is also below $S_C$
for negative energies,
it continues to rise for positive energies.
$S_G$
goes to the value
$k_B N \ln 2$
at the highest energy,
rather than the value of 
$k_B \ln 2$
that would correspond to the 
two-fold degeneracy of the highest energy level.
The negative slope of $S_C$ and $S_B$ for positive energies 
is consistent with this model exhibiting negative temperatures,
but $S_G$ predicts only positive temperatures.

The  expressions for the entropy
seen in
Fig.~\ref{Ising_plot} 
for the Ising model 
are qualitatively the same as those seen 
in Fig.~1
of Ref.~\cite{RHS_continuous}
for  independent two-level objects.
Equilibrium between two Ising models of different sizes 
would also have the same behavior 
as that shown 
in Fig.~2
of Ref.~\cite{RHS_continuous},
which demonstrated that 
$S_G$ cannot predict the equipartition of energy correctly.
The reason is again due to the size dependence 
of $T_G$ for positive energies.
For a given positive energy per spin,
the larger system would have a larger value 
of $T_G= 1/(\Delta S_G / \Delta E)$,
so that the smaller system would need more energy to match it.
Since this would violate equipartition of energy,
the predictions of $S_G$ are incorrect.

\subsection{Quantum simple harmonic oscillators}
\label{subsection: Q SHO}

In Ref.~\cite{RHS_continuous},
exact results were obtained for 
 $N$ quantum simple harmonic oscillators,
for which 
$S_B$ and $S_G$ 
are both  defined on a discrete set of energies.
They both lie below the canonical entropy 
for any finite $N$,
with $S_G$  above $S_B$.
In the limit 
$N \rightarrow \infty$,				
both $S_B$  and $S_G$ 
are defined for continuous energies 
and agree with the canonical entropy.
Although the system is obviously extensive
because it is composed of identical, independent
components,
 neither $S_B$  nor $S_G$ 
is exactly extensive for finite systems.
The canonical entropy, $S_C$, is extensive.
 We refer to 
  Ref.~\cite{RHS_continuous}
for a plot of the behavior.

\section{First-order phase transitions}
\label{section: first order}

In this section,
 we analyze first-order phase  transitions for various conditions.
We start  with the first-order phase transition 
of the  twelve-state Potts model,
which has short-range interactions.
Then we show that the thermodynamic limit 
obscures the physics of this transition
by hiding the terms responsible for it.
We then analyze the behavior of systems with long-range interactions,
showing that it is essentially the same,
 even though the appearance 
of the 
Boltzmann 
entropy per site in the thermodynamic limit
is quite different.

\subsection{The twelve-state Potts model}
\label{subsection: Potts}

\begin{figure}[htbp]
\begin{center}
 \includegraphics[width=\columnwidth]{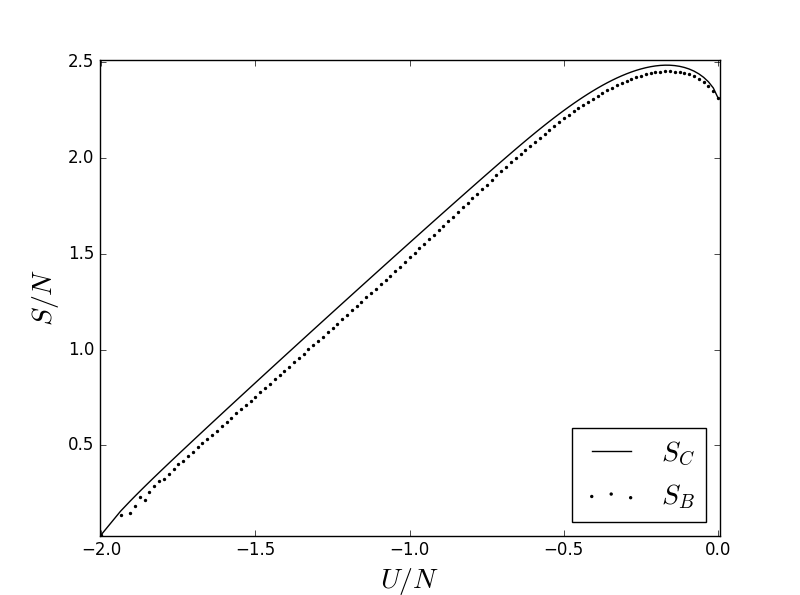}\\
  \caption{This plot shows 
  $S_B/N$  
  (the logarithm of 
  the density of states divided by $N$)
    for a two-dimensional
    twelve-state Potts model 
  on a $8 \times 8$ lattice
  obtained from a 
  Wang-Landau simulation\cite{Wang_Landau},
  along with 
  the canonical entropy per site, $S_C(U)/N$
calculated from that density of states.
  }
  \label{Potts_q12_L8}
\end{center}
\end{figure}

The two-dimensional Potts model with 
$q>4$ states 
is known to exhibit a first-order phase transition.
The Hamiltonian of the model can be written as 
\begin{equation}
H_{P} = - J \sum_{\langle j,k \rangle}   \delta_{\sigma_j , \sigma_k}  \,     ,
\end{equation}
where 
$ \delta_{a,b}$ is the Kronecker delta,
and the sum is over nearest-neighbor pairs of sites.  
We simulated the $q=12$
model with the 
Wang-Landau algorithm 
to obtain estimates of the density of states
$\omega_P (E_n)$\cite{Wang_Landau}.
The results of the simulation are shown in
Fig.~\ref{Potts_q12_L8}
for both the Boltzmann entropy
$S_{B}$
and the canonical entropy
$S_{C}$.
As for models discussed above,
$S_{C}$ lies above 
$S_{B}$ for all energies except the ground state energy,
for which they both have the value 
$k_B \ln 12$,
and for the maximum energy, where they also agree.


\begin{figure}[htbp]
\begin{center}
 \includegraphics[width=\columnwidth]{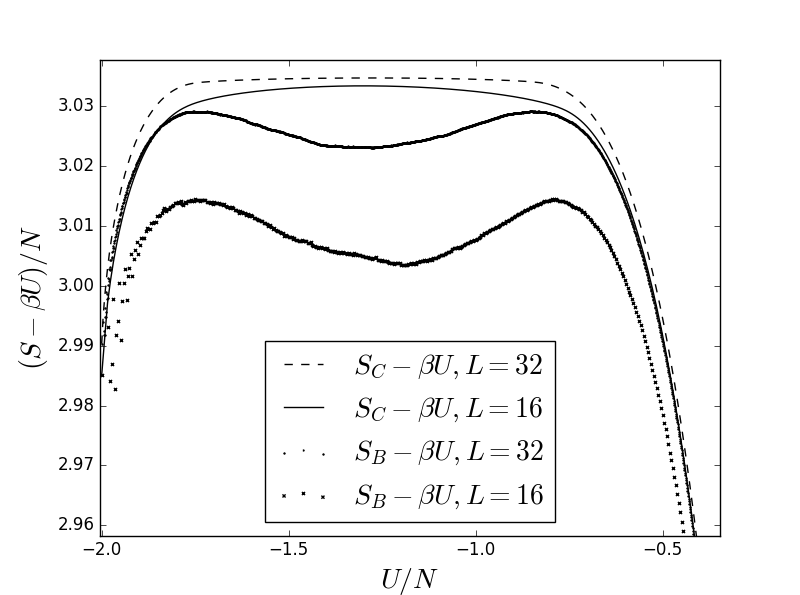}\\
  \caption{This plot compares 
  $\left(S(U) - \beta U\right)/N$ 
  for the canonical entropy, $S_C$,
  and the Boltzmann entropy, $S_B$.
The data is
 for   a  two-dimensional, 
 twelve-state  Potts model 
  on 
   $16 \times 16$ 
and
   $32 \times 32$ 
  lattices.
  The $L=16$ curves are  the second from the top ($S_C$) and at the bottom ($S_B$).
  For $L=32$ the curve for   $S_C$
is at the top, and  $S_B$ is second from the bottom.
  As in Fig.~\ref{Potts_q12_L8},
  the data was
  obtained from  
  Wang-Landau simulations\cite{Wang_Landau}.
  The exact value of the inverse critical temperature,
 $\beta_c =1.49607$.				
  For the 
     $16 \times 16$ 
lattice, $\beta = 1.48917$ ($0.46\%$ difference from $\beta_c$) was used to create the plots,
and
  for the 
     $32 \times 32$ 
     lattice, $\beta = 1.49492$ ($0.077\%$ difference from $\beta_c$) was used.
  }
  \label{Potts_16_32}
\end{center}
\end{figure}

Since the details of the first-order transition 
are difficult to see in 
Fig.~\ref{Potts_q12_L8},
we have replotted the data 
differently
in 
Fig.~\ref{Potts_16_32} 
for 
$L \times L$ lattices,
with
$L=16$ and $L=32$.
This plot shows 
   $( S(U) - \beta_c U)/N$ 
for both $S_{C}$ and $S_{B}$. 
The dip in $S_{B}$ 
with positive curvature
contrasts with 
$S_{C}$,
which  has
negative curvature
everywhere,
as required by thermodynamic stability.
The flattening of the curve for 
$S_{C}$ is just visible
between $L=16$ and $L=32$.
Note that 
because of the negative curvature of $S_{C}$,
a double-tangent construction 
would not produce the correct entropy
for any finite $L$.


\subsection{First-order transitions and the approximation of an infinite system}
\label{subsection: first order and thermodynamic limit}

Examining Fig.~\ref{Potts_16_32},
we see that the difference between $S_B$ and $S_C$
is smaller for the $32\times 32$ data than it is for 
$16 \times 16$.
This is a general property of 
first-order transitions in 
systems with short-range interactions,
when the entropy is divided by $N$.
The coexistence states are suppressed 
by the boundary energy between the two phases.
If $L$ is the linear dimension of the system,
then the interfacial free energy 
scales as 
$L^{d-1}$
or
$N^{1-1/d}$, 
where   
$d$ is the dimension.
In terms of the entropy per site,
these terms are of order
$N^{-1/d}$,
so that when the entropy per site is calculated 
for an infinite system,
their (apparent) contribution disappears.
The 
usual representation of the
thermodynamic limit gives 
the impression is  that the 
entropy contribution of these coexistence states
has vanished\cite{Touchette_introduction_2004,PhysRevE.74.010105,Touchette_equivalence_2011,Touchette_general_equivalence_2011}.

However,
the effects of the interfacial terms do not vanish.
The coexistence states are suppressed 
by interfacial energy terms 
that \textit{diverge} as 
$N^{1-1/d}$
(or $L$ for this model).    
As a result,
the Boltzmann entropy is not concave in the coexistence 
region for any finite system,
which  violates  a stability criterion.
That violation becomes infinitely strong
in 
an infinite system.
If this system is put into thermal contact 
with any other system,
it will leave the eigenstate for a 
distribution 
with a much higher probability.  

As Fig.~4 shows,
the actual behavior of the entropy for large systems is complicated.
Outside the region of the first-order transition,
$S_B-S_C$ scales as $L^{1/4}$.
Since the data is plotted as 
$\left(S_B-S_C  \right) L^{-1/4}$ vs. $U/N$,
the curves for 
$L=8$, $16$,  and  $32$
agree outside the first-order region.

In the first-order region,
$S_B-S_C$ is expected to scale as $L$,
which is the length of the interface between the 
ordered and disordered phases.
The data is consistent with that behavior,
but is still far from the asymptotic region.
A plot of 
$\left(S_B-S_C  \right) /L$ vs. $U/N$
for very large $N$
should show curves that go to zero
(as $L^{-3/4}$)
outside the first-order region,
but go to  a non-zero function
in the thermodynamic limit
inside the first-order region.


\begin{figure}[htbp]
\begin{center}
 \includegraphics[width=\columnwidth]{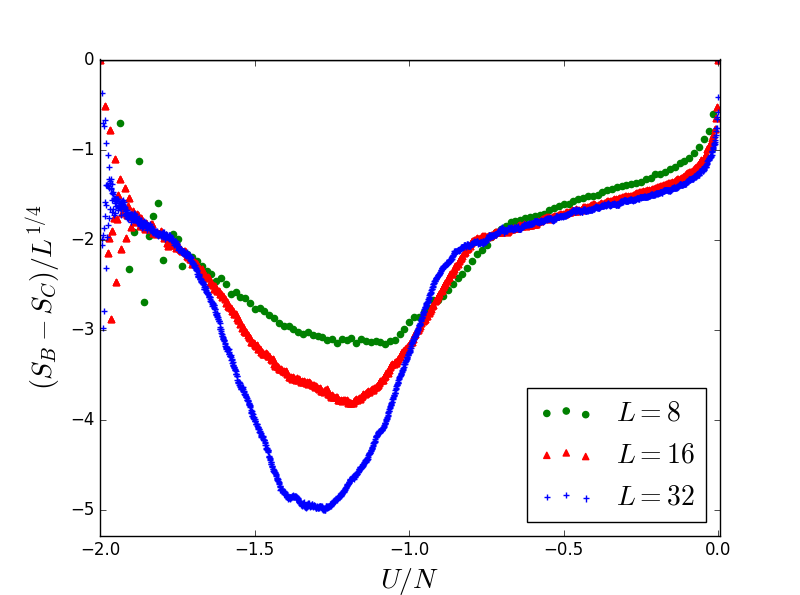}\\
  \caption{This plot shows
    $\left(S_B-S_C\right)/L^{1/4}$
  as a function of  $U/N$
   for   a  two-dimensional 
 twelve-state  Potts model 
  on an
  $8 \times 8$,
   $16 \times 16$, 
and
   $32 \times 32$ 
  lattices.
  As in Fig.~\ref{Potts_q12_L8} and \ref{Potts_16_32},
  the data was 
  obtained from  
  Wang-Landau simulations\cite{Wang_Landau}.
  As seen from the plot,
the region outside the first-order transition 
scales with $L^{1/4}$.
The region inside the first-order region
grow more rapidly.
It is expected to scale with $L$,
which is the length of the interface.
The data are consistent with this expectation,
but is too far from the asymptotic behavior to confirm it.
  In this representation,
  $S_B$ does not become identical with $S_C$ in the thermodynamic limit.
  }
  \label{Potts_16_32_rescale}
\end{center}
\end{figure}

This is not the usual way of representing 
the thermodynamic limit,
but it does display the information about the entropy.
A similar plot would exhibit the non-extensive part of the entropy
in	a gas in a container with walls that adsorbed particles,
as mentioned in the introduction.

For a system with long-range interactions,
the ``interface'' between the two phases 
is of the same dimension
as the system,
making the interfacial terms go as 
$N$ instead of $N^{1-1/d}$.
This makes no essential difference 
in the behavior of the first-order transition,
but it \emph{seems}
to make a difference 
because it affects the entropy per site (or per particle)
for an infinite system
 (thermodynamic limit).
For long-range interactions,
the contributions of the interface
 that cause the first-order transition
do not vanish in the usual representation 
of the thermodynamic limit.

Because of the different appearance of the first-order transition
in the presence of long-range interactions
for an infinite system,
we give 
an explicit analysis of 
a generic quantum model in the next subsection.
However, it is important to remember that 
a first-order transition 
behaves in essentially  the same way,
with or without long-range interactions --
only its representation is different.
As we will show,
the thermodynamic entropy $S_C$
is qualitatively the same in both cases.

\subsection{A generic model  with
long-range interactions and 
 a first-order transition  }
\label{subsection: generic first order}

We now introduce a Gaussian dip in the 
density of states of  a  generic quantum  model
to produce a first-order phase transition.
Our  model density of states is
\begin{eqnarray}\label{First_order_DoS_1}
\ln \omega_1(E_j) 
&=&
 A N \left( \frac{E_j}{N} \right)^{\alpha}    
 + B  N \exp \left( - \frac{ E_{N,0}^2 }{ 2 \sigma^2_N } \right)     \nonumber \\
&& - B  N \exp \left( - \frac{ ( E_j - E_{N,0} )^2 }{ 2 \sigma^2_N } \right)    .
\end{eqnarray}
The center of the Gaussian term is taken to be 
\begin{equation}\label{center f}
E_{N,0} = f  N \epsilon,
\end{equation}
and the width of the Gaussian is  
\begin{equation}\label{width g}
\sigma_N = g E_{N,0} .
\end{equation}
For our examples, 
we  take 
$\epsilon =1$,
$\alpha=0.5$,
$A=1$, $B=0.4$, $f=2$, and $g=0.1$,
but other values give similar results.

The partition function is  given by 
\begin{equation}
Z_N = \sum_j  \exp ( - \beta E_j )  \, \omega (E_j) ,
\end{equation}
and the thermodynamic energy is
\begin{equation}
U_N =   \frac{1}{Z_N}
\sum_j   E_j  \exp ( - \beta E_j ) \, \omega(E_j)    .
\end{equation}

\begin{figure}[htbp]
\begin{center}
 \includegraphics[width=\columnwidth]{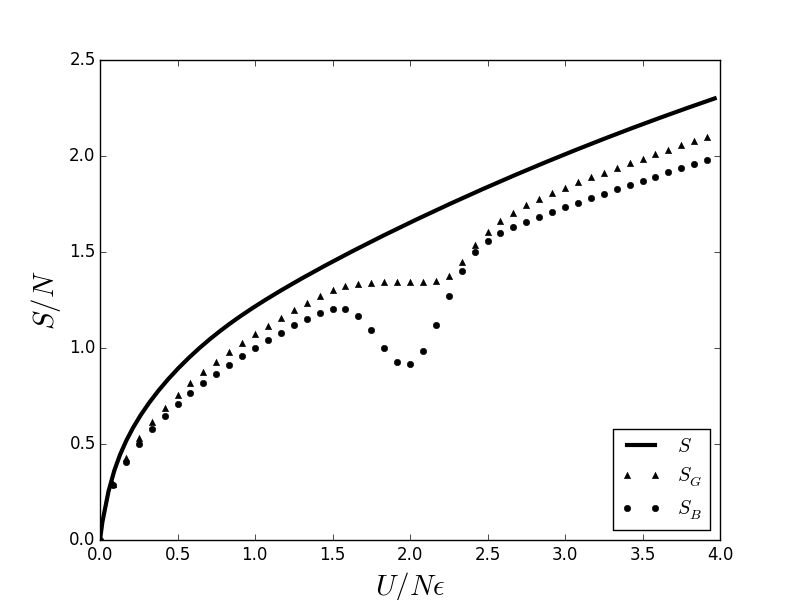}\\
  \caption{The plot shows 
  $S_C$, $S_G$, and $S_B$
  for the generic model of a first-order transition 
  as given in 
  Eq.~(\ref{First_order_DoS_1}),
  with $N=12$.
  While $S_G$ differs from $S_B$,
  it also shows a region with a positive second derivative.
  The canonical entropy $S_C$
  has a negative second derivative at all points.
    }
  \label{Single_S_U_plot}
\end{center}
\end{figure}

Fig.~\ref{Single_S_U_plot}
 shows $S_C$, $S_B$, and $S_G$
as functions of the energy $U_N$
for $N=12$.
The dip in the density of states $\omega(E)$
is reflected in $S_B$,
which is the logarithm of $\omega(E)$.
A modified dip is also seen in 
$S_G$;
the first derivative is non-negative,
but the second derivative has a positive region.
The canonical entropy 
$S_C$
shows no dip.
It has a negative second derivative 
at all points,
as required by thermodynamic stability.


\begin{figure}[htbp]
\begin{center}
 \includegraphics[width=\columnwidth]{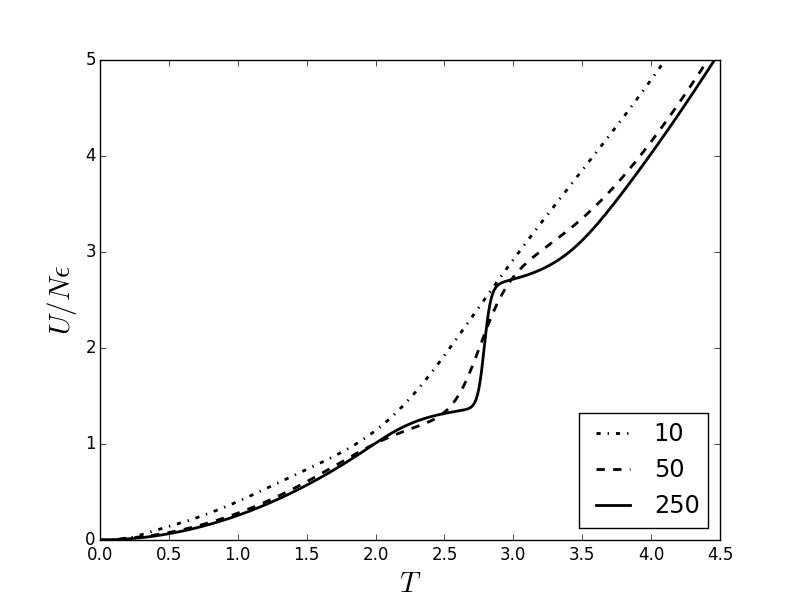}\\
  \caption{The plot shows the dimensionless energy,
  $U/N \epsilon$,
  as a function of temperature
  for the generic model of a first-order transition 
  as given in 
  Eq.~(\ref{First_order_DoS_1}).  
  The dip in the density of states has the form of a negative Gaussian.
  Computations were  made 
  for $N=10, 50$, and $250$.
  The constants $A$, $B$, $f$, and $g$
are taken to be $1.0$, $0.4$, $2.0$,  and $0.25$.
    }
  \label{First_S_U_T_plot}
\end{center}
\end{figure}

The development of the
 first-order transition can  be seen 
in Fig.~\ref{First_S_U_T_plot}.
This plot
 shows 
 a discontinuity forming
 in the thermodynamic energy 
$U$
as a function of temperature
$T$  for
$N=10, 50$, and $250$.


\begin{figure}[htbp]
\begin{center}
 \includegraphics[width=\columnwidth]{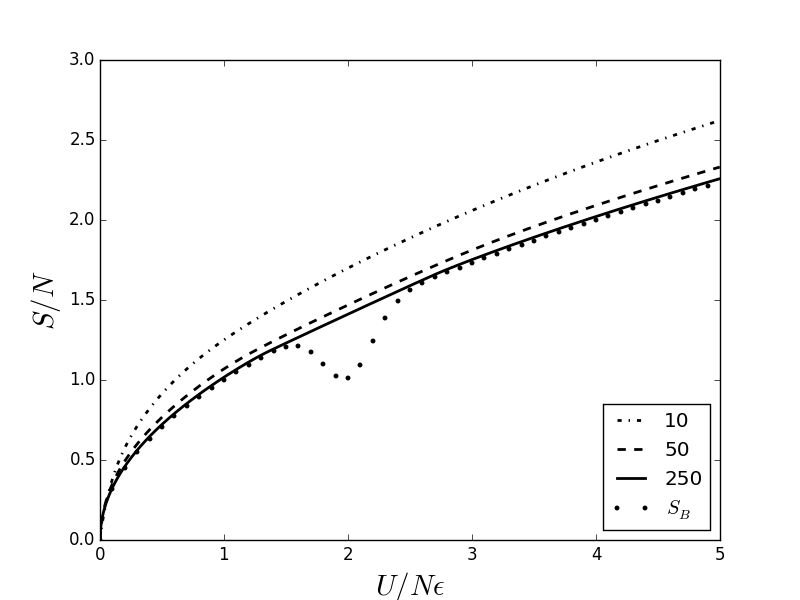}\\
  \caption{The solid line is a plot of the
  canonical entropy 
  $S_C$
   of the generic model of a first-order transition 
  with a Gaussian dip
  for $N = 10$ (dash-dot), $N=50$ (dashes),  and $250$ (solid line)
  as a function of the dimensionless thermodynamic energy $U/N \epsilon$,
  as given in 
  Eq.~(\ref{First_order_DoS_1}).
  The small circles show $S_B$
 for $N=10$.
   The constants $A$, $B$, $f$, and $g$
are taken to be $1.0$, $0.4$, $2.0$,  and $0.25$.
    }
  \label{First_S_omega_plot}
\end{center}
\end{figure}


Fig.~\ref{First_S_omega_plot}
shows a scaled plot of the 
canonical entropy $S_C$
as a function of energy 
for $N = 10, 50, $ and $250$.
Values of 
$S_B$
are only shown for $N=10$,
but 
$S_B$
for other values of $N$ 
also  follow the  curve
given in 
Eq.~(\ref{First_order_DoS_1}).
The dip in $S_B$ (and $S_G$)
due to the second term in 
Eq.~(\ref{First_order_DoS_1})
for $\ln \omega_1$
remains for all $N$,
creating a region with a positive 
second derivative with respect to $U$.
Stability requires that
the   entropy must have a non-positive 
second derivative 
for all $N$ and all energies,
even for long-range interactions.
In the limit
$N \rightarrow \infty$,
the canonical entropy develops a straight region that 
corresponds to the first-order transition,
just as it does in the case of short-ranged interactions.

\section{Classical models}
\label{section: classical models}

There are several classical 
models
with continuous 
densities of states that 
 provide interesting examples for comparisons
 between 
the canonical entropy 
$S_C$
and the microcanonical entropies $S_B$ and $S_G$.
We  begin with a simple power-law density of states
to demonstrate the method.
We then 
compare the definitions for a more complicated model suggested by 
Hilbert, H\"anggi, and Dunkel\cite{HHD_2014}.
To clarify the features of the model,
we consider three cases for which we can obtain analytic results:
an oscillating density of states,
and  both an 
exponentially
increasing and a decreasing  
density of states.
All models in this section have densities of states 
defined only for non-negative values of the energy.

\subsection{A power-law density of states}
\label{subsection: constant DoS}

A simple classical model density of states is given by a power law.
\begin{equation}\label{power law DoS}
\omega(E) = A E^n
\end{equation}
The partition function
for Eq.~(\ref{power law DoS}) 
 is found by integration of $\omega(E)$to be
\begin{equation}
Z
=
\int_0^{\infty}
\exp( - \beta E )
\omega(E) dE
=
A
 \beta^{-1-n} 
n!     \,  ,
\end{equation}
and the thermodynamic energy is
\begin{equation}
U = - 
\frac{ \partial \ln Z }{ \partial \beta }      
=
(n+1)/ \beta 
=
(n+1) k_B T         .
\end{equation}

Note that  
 a system of 
 $N=n+1$ simple harmonic oscillators
with frequency 
$\tilde{\omega}$
has a power-law density of states
with 
$A=  \left( \hbar \omega \right)^{-N} / (N-1)!  $. 
Its partition function   is
$Z_{SHO} = \left( \beta \hbar \tilde{\omega}\right)^{-N}$,
with $U=Nk_B T$.
A classical ideal gas also has a density of states of this form,
with $n=3N/2-1$
and an appropriate choice of $A$.

The canonical entropy is then
\begin{equation}
S_C
=
k_B 
\left[
(n+1)
\ln U
+
B_n
\right]       ,
\end{equation}
where
\begin{equation}
B_n
=
\ln A + \ln (n!) - (n+1) \ln \left( n+1  \right) 
+
(n+1)
\end{equation}
Note that as 
$T \rightarrow 0$,
$U \rightarrow 0$,
and 
 $S_C \rightarrow - \infty$,
which is a general classical result.
Naturally, 
this behavior violates the third law of thermodynamics,
which is only true for quantum systems.

\begin{figure}[htbp]
\begin{center}
 \includegraphics[width=\columnwidth]{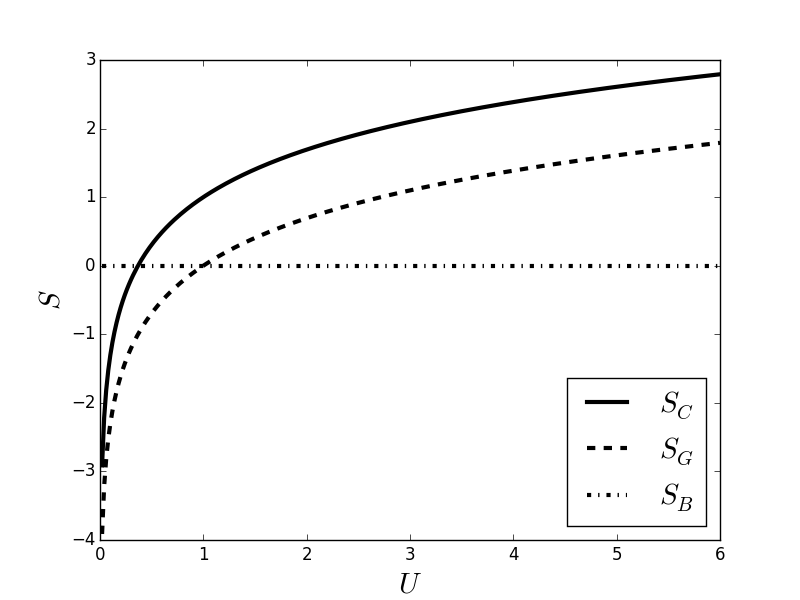}\\
  \caption{This plot shows 
  the canonical entropy 
  $S_C$
  of the 
  power-law  model density of states given in 
  Eq.~(\ref{power law DoS})
  with the exponent $n=0$,
  along with the corresponding 
  curves for 
  $S_B$ and $S_G$. 
  The constant value of the density of states 
  has been chosen to be $A=1$.
  }
  \label{constant_DoS}
\end{center}
\end{figure}

The corresponding expressions for 
the microcanonical entropies are
\begin{equation}
S_B = k_B \ln ( A \epsilon U^n)= k_B \left[ n \ln U + \ln ( A \epsilon ) \right] ,
\end{equation}
where $\epsilon$ is a constant with units of energy, 
and 
\begin{equation}
S_G = k_B \ln ( A \epsilon U^{n+1})= k_B \left[ (n+1) \ln U + \ln ( A \epsilon ) \right] ,
\end{equation}
In the limit of large $n$,
the three expressions for the entropy agree.
For small values of $n$, the expressions for $S_C$ and $S_G$
differ only by a constant,
which has no significance in classical statistical mechanics.
The differences with $S_B$ are greatest for $n=0$,
which can be seen from 
Fig.~\ref{constant_DoS}. 
which shows plots of 
all three expressions for entropy.
The Boltzmann entropy $S_B$ is
 wrong in this extreme case,
 since it is simply a constant,
 but it is correct for large values of $n$.


\subsection{The Hilbert, H\"anggi, and Dunkel  model}
\label{subsection: HHD model}

Hilbert, H\"anggi, and Dunkel (HHD)
have introduced an interesting  model density of states
in Ref.~\cite{HHD_2014}.
In their Eq.~(23),
they gave the integral of their density of states as
\begin{equation}\label{HHD_Omega}
\Omega(E)
=
\frac{ 2 E}{\epsilon}  
+
 \exp \left[  \frac{E}{ 2 \epsilon}  
- \frac{1}{4} \sin \left( \frac{ 2 E }{\epsilon} \right)
\right]       .
\end{equation}
The density of states itself can be  found from the derivative
\begin{equation}\label{omega_Omega}
\omega(E) = \frac{ \partial  \Omega(E) }{ \partial E }  ,
\end{equation}
and the explicit form for $\omega$ is 
\begin{eqnarray}\label{omega_2}
\omega(E)
&=&
 \frac{1}{ 2 \epsilon} 
\left(
 1
- \cos \left( \frac{ 2 E }{\epsilon} \right)
\right)          
  \exp \left[  \frac{E}{ 2 \epsilon}  
- \frac{1}{4} \sin \left( \frac{ 2 E }{\epsilon} \right)    \right]        \nonumber \\        
&&
+ 2 / \epsilon   .
\end{eqnarray}

There is a problem with the form of 
$\Omega(E)$
in
Eq.~(\ref{HHD_Omega})
because 
$\lim\limits_{E \rightarrow 0 } \Omega(E)= 1$,
although the limit should equal zero.
The expression for 
$\Omega(E)$ consistent with 
$\omega(E)$  is
\begin{equation}\label{HHD_Omega_corrected}
\Omega_{\textrm{corrected}}(E)
=
\frac{ 2 E}{\epsilon}  
+
 \exp \left[  \frac{E}{ 2 \epsilon}  
- \frac{1}{4} \sin \left( \frac{ 2 E }{\epsilon} \right)
\right]      
-1 .
\end{equation}
Nevertheless,
 we performed our calculations 
with  
Eqs.~(\ref{HHD_Omega})
and
(\ref{omega_2})
to facilitate comparison with 
Ref.~\cite{HHD_2014}.

\begin{figure}[htbp]
\begin{center}
 \includegraphics[width=\columnwidth]{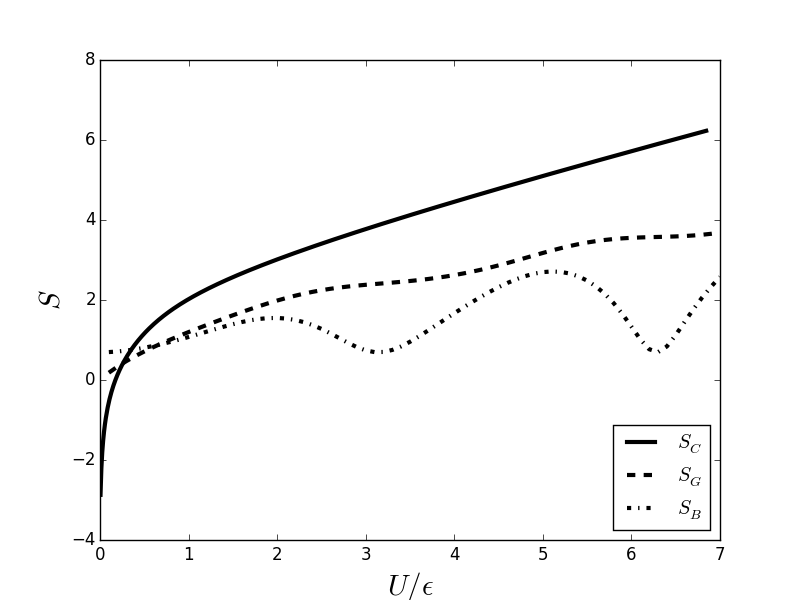}\\
  \caption{This figure shows the 
  canonical entropy $S_C$  vs. the energy $U$
  in comparison with both 
  $S_G$ and $S_B$
  for the 
  HHD model 
  density of states introduced 
  in Ref.~\cite{HHD_2014}.
  Eqs.~(\ref{HHD_Omega})
and
(\ref{omega_2})
were used 
rather than the corrected equation for $\Omega$
in 
Eq.~(\ref{HHD_Omega_corrected}),
to facilitate comparisons with 
Fig.~1.b in Ref.~\cite{HHD_2014},.
  While $S_B$ shows strong oscillations,
  which are mirrored in the repeated flat regions of $S_G$,
  $S_C$ shows a smooth increase at all energies.
  At high energies, the slope of 
  $S_C$ vs. $U$ goes to a constant value,
  as expected
  from comparisons with the entropy 
  of the exponential model shown in 
Fig.~\ref{Exponential_DoS}.
}
    \label{PLOT_HHD_S}
\end{center}
\end{figure}

Fig.~\ref{PLOT_HHD_S}
shows 
$S_C$, 
which is computed numerically,
$S_G=k_B \ln \Omega(U)$, and $S_B=k_B \ln \omega(U)$
for the HHD model.
The curves for 
$S_G$ and $S_B$
agree with those given
in  Fig.~1.b of 
Ref.~\cite{HHD_2014}.
$S_B$ has an infinite sequence of oscillations,
with exponentially increasing magnitude.
Due to  
Eq.~(\ref{omega_Omega}),
the minima in 
$S_B$
correspond to minima in the slope of 
$S_G$.
In contrast,
the plot of 
$S_C$ vs. $U$
is remarkably smooth,
with none of the oscillations 
that appear in 
$S_B$ and $S_G$.
As $U \rightarrow 0$,
we again see that  
$S_C \rightarrow -\infty$,
which is typical for the entropy 
of a classical system.
Neither 
$S_B$ nor  $S_G$ exhibits 
this classical divergence,
although the corrected version of 
$\Omega$ in 
Eq.~(\ref{HHD_Omega_corrected})
would show this  divergence.


\begin{figure}[htbp]
\begin{center}
 \includegraphics[width=\columnwidth]{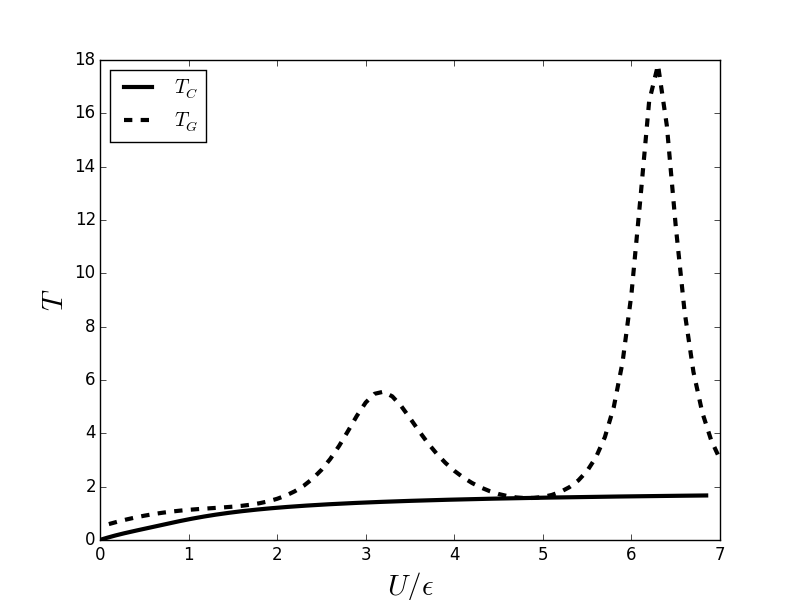}\\
  \caption{This figure shows the 
  canonical  temperature $T_C$  vs. the energy $U$
  in comparison with  the Gibbs temperature 
  $T_G$
    for the density of states introduced 
  in Ref.~\cite{HHD_2014}.
  While $T_G$ shows repeated peaks,
    $T_C$ shows a smooth increase at all energies.
  At large  energies, 
  $T_C$ goes to a constant value, $T_{\textrm{max}}$,
  which is the maximum temperature
  for this model.
    }
  \label{PLOT_HHD_T}
\end{center}
\end{figure}

For large values of the energy $U$,
the slope of $S_C$ 
goes to a constant,
which is the inverse of the maximum temperature of the model.
The unusual phenomenon of a maximum temperature 
 is due to the exponentially increasing 
part of 
Eq.~(\ref{HHD_Omega}).
The energy dependence of the canonical  temperature 
$T_C=1/(\partial S_C / \partial U)$
is shown 
as a function of the energy in 
Fig.~\ref{PLOT_HHD_T}.
To see how a maximum temperature  arises,
we investigate a simpler  model 
with an exponentially increasing density of states
below  in 
Subsection~\ref{subsection: increasing exponential}.


Fig.~\ref{PLOT_HHD_T}
 also shows 
 the Gibbs temperature $T_G$,
and is consistent with 
 Fig.~1.c
 of 
 Ref.~\cite{HHD_2014}. 
Since 
\begin{equation}
\frac{1}{T_G }
= 
\frac{\partial S_G }{ \partial E } 
=
\frac{ \omega (E) }{ \Omega (E) }   ,
\end{equation}
Fig.~\ref{PLOT_HHD_T}
shows a peak in $T_G$
corresponding to  each minimum  of 
$\omega(E)$.
These peaks in $T_G$
mean that a given value of the temperature $T_G$ 
corresponds to a sequence of values of $U$.
This has been noted 
by HHD,
who regard it as correct thermodynamic behavior,
 writing that,
``the same temperature value $T_G$ or
$T_B$ can correspond to vastly different energy values \cite{HHD_2014}.''

We believe that a multiplicity of thermodynamic energies 
for a single temperature
is non-physical.
In support of this position,
there is a physically reasonable 
one-to-one relationship between the canonical temperature 
$T_C$
and the energy $U$.

The sequence of peaks in $T_G$
as a function of $U$
is investigated below 
in 
Subsection \ref{subsection: simple oscillate}
for a simpler model
with a density of states
that oscillates, but does not diverge exponentially
for large energies.

\subsection{An oscillating density of states}
\label{subsection: simple oscillate}

To investigate the predicted sequence of peaks 
in $T_G$ as a function of $U$,
discussed  in 
Subsection \ref{subsection: HHD model},
we introduce 
a simpler model
with an oscillating   density of states,
for which the canonical entropy that can be obtained analytically.
\begin{equation}\label{Oscillating_DoS}
\omega(E)
=
A +B - B  \cos \left(  c E \right)
\end{equation}
The Boltzmann and Gibbs  entropies are then
\begin{equation}
S_B
=
k_B
\ln \omega(E) 
=
k_B
\ln  \left[
A +B
-
B   \cos ( c E )
 \right]     
\end{equation}
and
\begin{equation}
S_G
=
k_B \ln
\Omega(E)
=
k_B \ln
\left[
(A +B)E
-
\frac{ B }{ c }
 \sin \left(  c E \right)
 \right]   .
\end{equation}

\begin{figure}[htbp]
\begin{center}
 \includegraphics[width=\columnwidth]{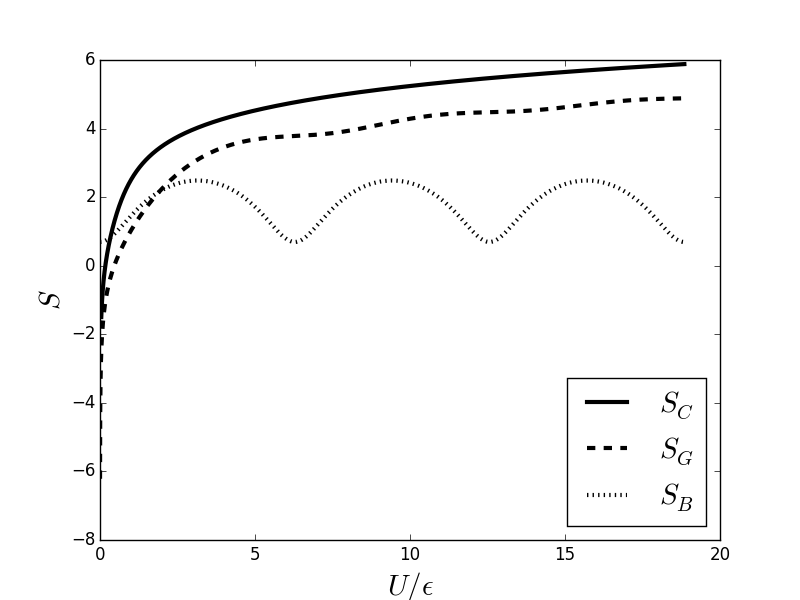}\\
  \caption{This plot compares the canonical entropy $S_C$
with the predictions of 
$S_B$ 
and 
$S_G$, 
for the oscillating density of states 
given in 
Eq.~(\ref{Oscillating_DoS}),
with parameters 
$a=1$ and $B=5$.
    }
  \label{PLOT_S_Oscillate}
\end{center}
\end{figure}

The partition function is
\begin{equation}
Z
=
\frac{A+B}{\beta} 
-
\frac{ B \beta  }{ c^2 + \beta^2}  ,
\end{equation}
 the energy is
\begin{eqnarray}
U
&=&
-
\frac{1}{Z}
\frac{ \partial Z }{ \partial  \beta}   \nonumber \\
&=&
\frac{1}{\beta}+
\frac{
2 B c^2 \beta
}{
( c^2 + \beta^2 )
\left( ( A+B) c^2 +A \beta^2 \right)  
}  .
\end{eqnarray}
and
 the canonical entropy is given by
$S_C=k_B  \ln Z  + \beta U$.
The expression for 
the canonical entropy
of this model
is rather  long,
but it easy to plot,
and
Fig.~\ref{PLOT_S_Oscillate}
shows 
$S_C$, $S_G$, and $S_B$. 
The oscillations in 
$S_B$ are reflected 
in oscillations of the slope in
$S_G$.
However, oscillations are entirely missing 
in the canonical entropy.

$S_C$ and $S_G$ 
show the typical low temperature behavior
of classical systems:
when 
$U \rightarrow 0$,
$T \rightarrow 0$,
and 
$S_C \rightarrow - \infty$.
As usual, 
$S_B$ 
does not have  this feature.


The predictions of  $S_B$ and $S_G$ 
for the behavior of the temperature
are similar to those for the HHD model discussed in 
Subsection~\ref{subsection: HHD model}.
The inverse temperature $\beta_B=1/k_B T_B$
is the derivative of 
$S_B$ with respect to $U$.
We have not plotted $T_B$ vs.~$U$,
but it can be seen from 
Fig.~\ref{PLOT_S_Oscillate}
that 
$S_B$
  oscillates between positive and negative slopes,
  with the slope 
vanishing at every maximum and minimum,
so that $T_B \rightarrow \pm \infty$
at these points.

\begin{figure}[htbp]
\begin{center}
 \includegraphics[width=\columnwidth]{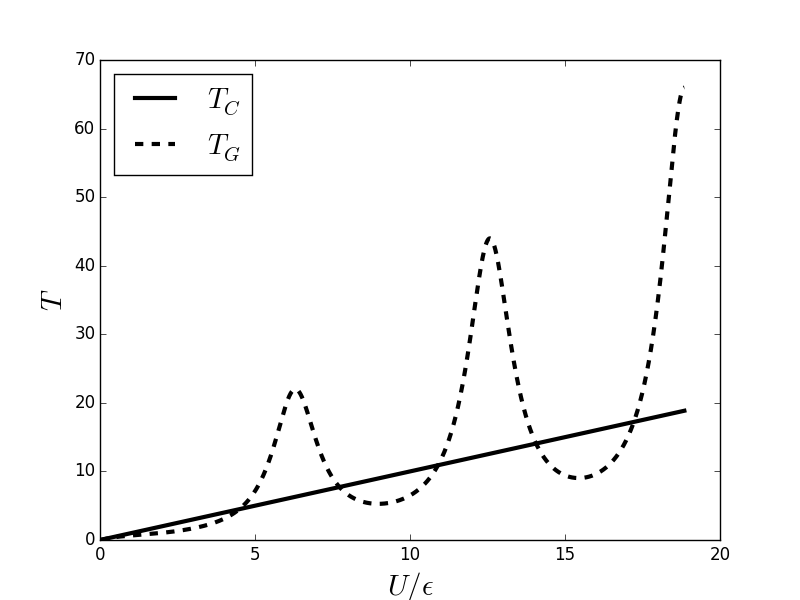}\\
  \caption{This plot compares
    $T_C$ derived from
   $S_C$,
with 
   the function $T_G$ derived from
$S_G$, 
for the oscillating density of states 
given in 
Eq.~(\ref{Oscillating_DoS}),
with parameters 
$a=1$ and $B=5$.
The peaks in the plot of $T_G$ mean that a given value of $T_G$  
can correspond to many different values of the energy $U$.
There is a single value of $U$ for any given value of $T_C$.
    }
  \label{PLOT_T_Oscillate}
\end{center}
\end{figure}

The behavior of 
$T_G$ as a function of $U$ 
is shown in 
Fig.~\ref{PLOT_T_Oscillate},
which shows a sequence of peaks corresponding to the 
minima in the slope of $S_G$ 
in 
Fig.~\ref{PLOT_S_Oscillate}.
This behavior is qualitatively 
the same as that seen in 
Fig.~\ref{PLOT_HHD_T}
(Subsection \ref{subsection: HHD model}),
and again
 has the  consequence that 
a given value of either 
$T_B$ or $T_G$ might correspond to 
many  values of the energy $U$.


\subsection{An increasing exponential density of states}
\label{subsection: increasing exponential}

An exponentially increasing density of states 
is  another   simplified version of 
the HHD model discussed in 
Subsection \ref{subsection: HHD model}.  
The simplified  model is
\begin{equation}\label{equation_exponential_DoS}
\omega( E ) = B  \exp( cE ) ,
\end{equation}
with $c>0$.

The partition function is  
\begin{equation}
Z
=
\frac{ B }{ \beta - c }     ,
\end{equation}
and the energy is
\begin{equation}
U 
=
\frac{
1
}{
  \beta - c
}
=
Z / B   .
\end{equation}
Note that the partition function $Z$  and the energy $U$ 
both diverge at 
$\beta_{\textrm{min}}=c$,
or 
$k_B T_{\textrm{max}}= 1/c$.
A peculiarity of this model,
which it shares with the HHD model 
(Subsection \ref{subsection: HHD model}),
 is that 
there is a maximum temperature,
and $U \rightarrow \infty$ as
$T \rightarrow T_{\textrm{max}}$.

The canonical entropy for the exponential density of states
as a function of $U$
 is given by
\begin{eqnarray}
S_C
&=&
k_B \left[ 
\ln \left(   BU \right)  
+  
 \beta U
 \right]       \nonumber \\
&=&
k_B \left[ 
\ln \left(   BU \right)  
+  
1 + c U
 \right]    .
 \end{eqnarray}

\begin{figure}[htbp]
\begin{center}
 \includegraphics[width=\columnwidth]{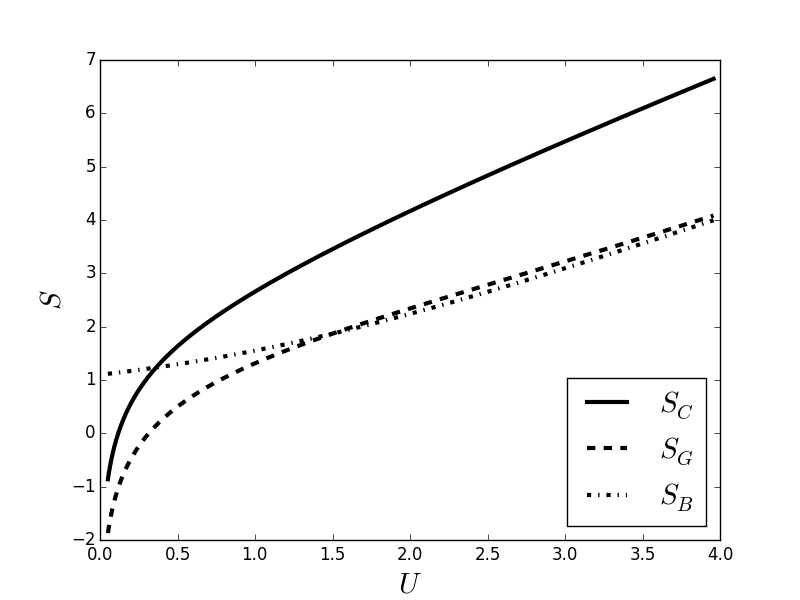}\\
  \caption{This plot shows 
  the canonical entropy 
  $S_C$
  of the 
  exponential  model density of states given in 
  Eq.~(\ref{equation_exponential_DoS}),
  along with the corresponding 
  curves for 
  $S_B$ and $S_G$.
  The parameters in 
  Eq.~(\ref{equation_exponential_DoS})
  were taken to be 
  $A=2$, and $B=c=\epsilon=1$.
  }
  \label{Exponential_DoS}
\end{center}
\end{figure}

The corresponding   expressions for 
$S_B$ and $S_G$ 
are
\begin{equation}
S_B = k_B \ln [  B  \epsilon \exp(cU) ]
=
 k_B [  \ln ( B  \epsilon) +  cU ]
\end{equation}
and 
\begin{equation}
S_G = k_B \ln [  (B/c) ( \exp(cU) - 1) ]
\end{equation}
All three expressions for entropy  are plotted in 
Fig.~\ref{Exponential_DoS}
as  functions of the energy.
Because the density of states increases so rapidly,
all three expressions agree for large $U$,
except for an additive constant 
that has no significance for classical statistical mechanics.
The slope of all three functions 
goes to the expected value of 
$\beta_{\textrm{min}}=c$ 
for large $U$.
Both $S_C$ and $S_G$ show the correct 
divergence to $-\infty$ as $U \rightarrow 0$,
but $S_B$ goes to a constant.


\subsection{An exponentially decreasing  density of states}
\label{subsection: decreasing exponential}

The case of 
an exponentially decreasing density of states 
turns out to be unexpectedly interesting.
We write
\begin{equation}\label{equation_exponential_DoS_decrease}
\omega( E ) = B \exp( - |c| E ) ,
\end{equation}
where we have indicated the negative argument 
of the exponential function explicitly 
to avoid confusion.

To calculate the canonical entropy,
we find that
the partition function is 
\begin{equation}
Z
=
\frac{ B }{ \beta  + |c| }  ,   
\end{equation}
the energy is
\begin{equation}
U
=
\frac{1 }{ \beta + |c|} ,
\end{equation}
so that  the two are related by  
\begin{equation}
Z = BU    .
\end{equation}
Note that  $Z$  and $U$ 
 diverge at 
$\beta_{\textrm{min}}=-|c|$,
but are well defined for all  temperatures
algebraically greater than $-1 /  k_B \vert c \vert$.

The canonical entropy
as a function of $U$
 is 
\begin{equation}
S_C
=
 k_B \left[ 
\ln \left(   B U   \right)  
+  
1
-
\vert c \vert  U   
 \right]  
 \end{equation}

The  expressions for 
$S_B$ and $S_G$ 
are  
\begin{equation}
S_B 
=
 k_B[   \ln  (B  \epsilon )  - |c|U ]
\end{equation}
and 
\begin{equation}
S_G = k_B \ln [  (B/|c|) ( 1 - \exp( - |c|U) ) ]
\end{equation}


\begin{figure}[htbp]
\begin{center}
 \includegraphics[width=\columnwidth]{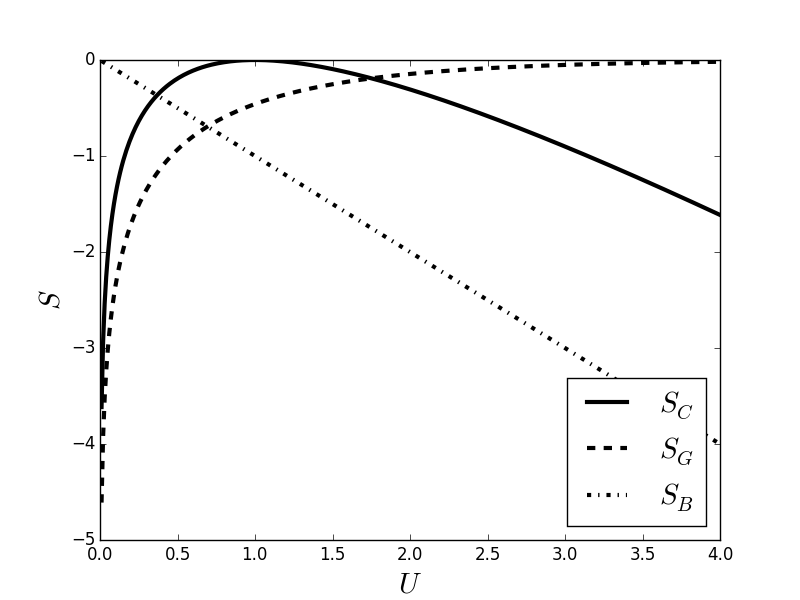}\\
  \caption{This plot shows 
  the canonical entropy 
  $S_C$
  of the 
 exponentially decreasing   
 density of states given in 
  Eq.~(\ref{equation_exponential_DoS_decrease}),
  along with the corresponding 
  curves for 
  $S_B$ and $S_G$.
  The parameters in 
  Eq.~(\ref{equation_exponential_DoS_decrease})
  were taken to be 
  $A=0$, and $B=\epsilon=1$,
  and $c=-1$.
  }
  \label{Exponential_DoS_decrease_S}
\end{center}
\end{figure}

Fig.~\ref{Exponential_DoS_decrease_S}
shows plots of 
$S_C$, $S_B$, and $S_G$
as functions of the energy $U$.
$S_B$ and $S_G$ contradict each other,
with $S_B$ having a constant negative slope
(implying the same negative temperature for all energies,
 while $S_G$ 
 has a positive slope and positive temperatures for all energies.
$S_C$ differs from both,
and we regard its predictions as correct.

\begin{figure}[htbp]
\begin{center}
 \includegraphics[width=\columnwidth]{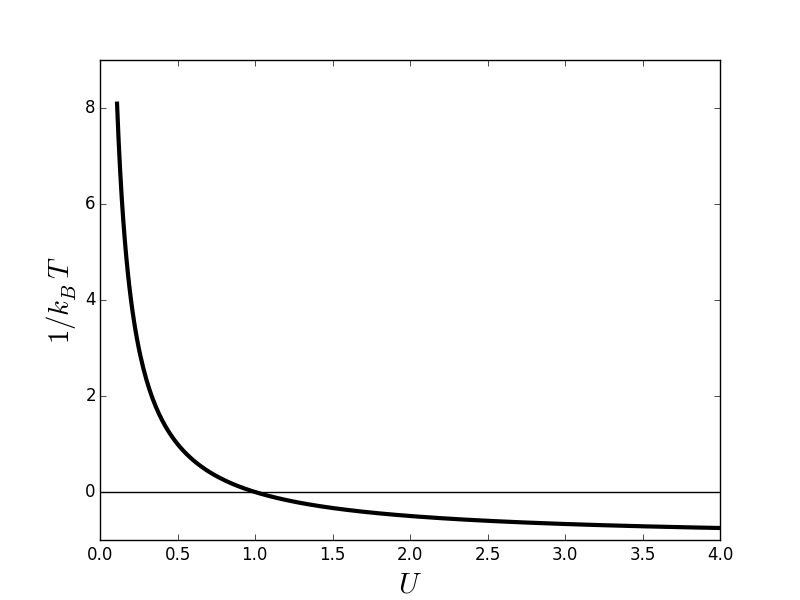}\\
  \caption{This plot shows 
  the  inverse canonical temperature 
  of the 
  exponentially decreasing 
  density of states given in 
  Eq.~(\ref{equation_exponential_DoS_decrease}),
  The parameters in 
  Eq.~(\ref{equation_exponential_DoS_decrease})
  were taken to be 
  $A=0$, and $B=\epsilon=1$,
  and $c=-1$.
  }
  \label{Exponential_DoS_decrease_T}
\end{center}
\end{figure}

The maximum  in the canonical entropy 
as a function of $U$ 
is somewhat surprising 
since the density of states is monotonically decreasing.
This maximum has the consequence that  the energy 
is finite 
for infinite temperature:
when $U = 1/|c|$,
$\beta_C = 0$ and $T_C=\infty$.
The canonical temperature 
$T_C$
is positive for 
$U < 1/ |c|$
and negative for 
$U > 1/ |c|$,
as shown in 
Fig.~\ref{Exponential_DoS_decrease_T}.
Such behavior is well-known for 
models with a maximum in the density of states,
like the Ising model. 
However, the density of states for this model 
is monotonically  decreasing.


\section{Conclusions}

We have shown that the three proposed definitions 
give very different expressions for the thermodynamic entropy 
for various models.
Since the thermodynamic entropy 
must be unique\cite{lieb1997guide},
only one of these definitions
of the entropy can be correct.

For the  models we have investigated,
 the canonical entropy,
 $S_C(U)$,
 gives physically reasonable predictions 
 for all thermodynamic properties.
 By construction,
 these predictions are identical to those 
 of the canonical ensemble,
 making entropy consistent with 
 Helmholtz free energy 
 and all other thermodynamic potentials.

For all quantum models discussed in 
Sections \ref{section: quantum without PT}
and 
\ref{section: first order},
neither 
$S_B$ nor $S_G$ 
gives the correct entropy 
for any finite system,
because they are defined 
only on a discrete set of energy values.
Only in  the limit 
$N \rightarrow \infty$,
$S_B$ and $S_G$ become 
continuous functions of $U$.
$S_B$  becomes quantitatively correct 
for the  models  discussed  in 
Section \ref{section: quantum without PT}. 
$S_G$ also agrees for values of the energy
corresponding to an increasing density of states,
but not for the decreasing density of states 
for positive energies 
in the independent-spin model
discussed in Ref.~\cite{RHS_continuous},
or in the Ising model,
shown in Fig.~\ref{Ising_plot}.
It is perhaps significant that no evidence has
been published to indicate that 
$S_G$ should be correct for a decreasing density of states.

In Section \ref{section: first order},
both $S_G$ and $S_B$  
are seen to disagree with the canonical entropy 
of the  twelve-state Potts model
for all system sizes.
They both exhibit a range of energies 
for which their curvature is positive,
which violates a condition of thermodynamic stability. 
If a system
 in an eigenstate
 in this region is brought into thermal contact 
 with an exact copy of itself,
 both systems will go to a new thermodynamic  state,
 and the equilibrium will be described by 
 the canonical  entropy.

In Fig.~\ref{Potts_16_32_rescale},
we have suggested a new way of representing $S_B$ 
in the thermodynamic limit,
for a system that exhibits a first-order transition.
The quantity 
$\left( S_B- S_C \right) / N^{1-1/d}$
does not vanish for $N \rightarrow \infty$
 in the region of a first-order phase transition,
so that the Bolzmann entropy and the canonical entropy
should not be regarded as equivalent.

The 
Gibbs and Boltzmann entropies for the 
generic density of states for a model with
long-range interactions and a 
first-order phase transition are also incorrect,
in that they both contain a region with a positive curvature,
even in the limit 
$N \rightarrow \infty$.
The canonical entropy shows no such anomaly. 

Our analysis 
has  shown that 
care must be taken when using the 
approximation 
of an infinite system
(thermodynamic limit)
for  first-order transitions.
When properly viewed,
first-order transitions are all essentially the same,
whether the forces are long-range or short-range.

For the classical (continuous) densities of states
discussed in Section \ref{section: classical models},
 $S_B$ and $S_G$ 
 only give good predictions for the power law density of states 
  for large values of $n$.
The Gibbs entropy correctly shows the classical divergence of $S_G$
for $U \rightarrow 0$
for all classical models except the HHD model 
discussed in 
Subsection \ref{subsection: HHD model}.
$S_G$
is essentially correct for the constant density of states 
discussed in Subsection \ref{subsection: constant DoS},
and the exponentially increasing density of states 
discussed in Subsection \ref{subsection: increasing exponential}.

Neither $S_G$ nor $S_B$ shows the correct behavior for the 
oscillating density of states in 
Subsections
\ref{subsection: HHD model}
and
 \ref{subsection: simple oscillate}.
The prediction of peaks in a plot of 
temperature as a function of energy
is a serious departure from physically reasonable behavior.

The case of a decreasing exponential density of states,
discussed in 
Subsection~\ref{subsection: decreasing exponential},
is particularly interesting in that the three definitions of entropy 
give completely different predictions for the thermodynamic behavior.
We regard the canonical prediction
of both positive and negative temperature regions as correct,
with high energies corresponding to the most negative temperatures.
This confirms the interpretation 
that negative temperatures are hotter than infinite temperature.

In summary,
the canonical entropy
provides a correct procedure for calculating  the thermodynamic  entropy
as a function of the energy $U$
for both quantum and classical systems
from the principles of statistical mechanics.
Although
the microcanonical definitions of entropy,
$S_B$ and $S_G$,
might give reasonable approximations
that are useful 
for a limited class of models,
they are not always correct,
and their predictions can be misleading.

 \section*{Acknowledgement}

RHS would like to thank
Jian-Sheng Wang
and
Oliver Penrose 
for their valuable comments,
and
Roberta Klatzky 
for many helpful discussions.
This research did not receive any specific grant from funding agencies in the public, commercial, or
not-for-profit sectors.

\makeatletter
\renewcommand\@biblabel[1]{#1. }
\makeatother

\bibliography{Entropy_citations_3}

\begin{thebibliography}{46}
\bibitem [{Bol()}]{Boltzmann_entropy}%
  \BibitemOpen
  \bibinfo {note} {Boltzmann derived the expression for the entropy for
  classical systems under the assumption that equilibrium corresponded to the
  maximum of the probability for two systems in equilibrium. The quantum
  version is due to Planck\cite{Planck_1901}. Planck is also responsible for
  the form of the entropy carved into Boltzmann's grave stone, $S=k \log W$
  (where ``W'' stands for the German word ``Wahrscheinlichkeit'', or
  probability), and the introduction of $k$ as the Boltzmann
  constant.}\BibitemShut {Stop}%
\bibitem [{Gib()}]{Gibbs_entropy}%
  \BibitemOpen
  \bibinfo {note} {We are using the term ``Gibbs entropy'' to refer to the
  definition of the entropy in terms of the logarithm of the volume of phase
  space with energy less than a given energy. The quantum version refers to the
  sum of all degeneracies of eigenstates with energies below a given energy. It
  is not to be confused with another definition of entropy due to Gibbs in
  terms of an integral of $\int \rho \ln \rho$, where $\rho$ is the probability
  of a microscopic state.}\BibitemShut {Stop}%
\end{thebibliography}

\end{document}